\documentclass[reprint,aps,prmaterials,twocolumn]{revtex4}

\usepackage[pdftex]{graphicx}
\usepackage{amsmath}

\begin{document}

\title{Ferroelectric enhancement of superconductivity in compressively strained SrTiO$_3$ films}

\author{Ryan Russell}
\thanks{These authors contributed equally to this work.}
\affiliation{Materials Department, University of California, Santa Barbara, California 93106, USA}

\author{Noah Ratcliff}
\thanks{These authors contributed equally to this work.}
\affiliation{Materials Department, University of California, Santa Barbara, California 93106, USA}

\author{Kaveh Ahadi}
\affiliation{Materials Department, University of California, Santa Barbara, California 93106, USA}

\author{Lianyang Dong}
\affiliation{Materials Department, University of California, Santa Barbara, California 93106, USA}

\author{Susanne Stemmer}
\affiliation{Materials Department, University of California, Santa Barbara, California 93106, USA}

\author{John W. Harter}
\email[Corresponding author: ]{harter@ucsb.edu}
\affiliation{Materials Department, University of California, Santa Barbara, California 93106, USA}

\date{\today}

\begin{abstract}
SrTiO$_3$ is an incipient ferroelectric on the verge of a polar instability, which is avoided at low temperatures by quantum fluctuations. Within this unusual quantum paraelectric phase, superconductivity persists despite extremely dilute carrier densities. Ferroelectric fluctuations have been suspected to play a role in the origin of superconductivity by contributing to electron pairing. To investigate this possibility, we used optical second harmonic generation to measure the doping and temperature dependence of the ferroelectric order parameter in compressively strained SrTiO$_3$ thin films. At low temperatures, we uncover a spontaneous out-of-plane ferroelectric polarization with an onset that correlates perfectly with normal-state electrical resistivity anomalies. These anomalies have previously been associated with an enhancement of the superconducting critical temperature in doped SrTiO$_3$ films, directly linking the ferroelectric and superconducting phases. We develop a long-range mean-field Ising model of the ferroelectric phase transition to interpret the data and extract the relevant energy scales in the system. Our results support a long-suspected connection between ferroelectricity and superconductivity in SrTiO$_3$, but call into question the role played by ferroelectric fluctuations.
\end{abstract}

\maketitle

\section{Introduction}

Despite its long history and contemporary interest, the superconducting phase of SrTiO$_3$ is still not well understood~\cite{collignon2019}. A central open question is how the superconductivity emerges from an exceptionally dilute Fermi gas. In this regime, the Fermi energy is less than the Debye frequency, making conventional Bardeen-Cooper-Schrieffer (BCS) and Migdal-Eliashberg theories of acoustic phonon mediated electron pairing inapplicable~\cite{takada1980}. It has become increasingly clear that an unconventional pairing mechanism is responsible for superconductivity in SrTiO$_3$. Given the proximity of the material to a polar instability, fluctuations of ferroelectric order are an obvious candidate~\cite{edge2015,dunnett2018,rowley2018}. Ferroelectricity and superconductivity, however, have a precarious relationship. Superconductivity requires free carriers, with the superconducting energy gap depending strongly on the density of states at the Fermi level, while free charges tend to rapidly screen electric dipole fields and prevent long-range spontaneous electric polarization. Indeed, only recently have long-sought examples of so-called ``polar metals'' been discovered~\cite{anderson1965,shi2013,filippetti2016,lei2018}. Because of this discordance, the fundamental relationship between ferroelectricity and superconductivity in SrTiO$_3$ is still a matter of debate.

The putative link between superconductivity and ferroelectricity has in recent years motivated studies of the effects of enhanced ferroelectric fluctuations on the superconducting properties of SrTiO$_3$. Experiments have reported an increase in the superconducting critical temperature $T_c$ by tuning the material towards a ferroelectric ground state using oxygen isotope substitution~\cite{stucky2016,tomioka2019} and alloying with calcium~\cite{rischau2017}. In thin films, epitaxial strain is also known to stabilize ferroelectricity~\cite{pertsev2000,haeni2004,verma2015,haislmaier2016}, and a striking recent study has shown that in-plane compressive strain can enhance the superconducting critical temperature by up to a factor of two~\cite{ahadi2019}. To establish a clear link between this superconducting enhancement and ferroelectricity, here we directly measure the ferroelectric order parameter of compressively strained films of electron-doped SrTiO$_3$. The metallic nature of these films precludes traditional electrical methods for measuring polarization. Optical second harmonic generation (SHG), on the other hand, is an all-optical technique that is acutely sensitive to inversion symmetry breaking, making it an ideal and sensitive probe of polar order~\cite{harter2015,harter2017,zhao2018}.

\section{Experimental Methods}

Samarium-doped SrTiO$_3$ films of thickness 200~nm were grown epitaxially on (001) (LaAlO$_3$)$_{0.3}$(Sr$_2$AlTaO$_6$)$_{0.7}$ (LSAT) substrates via hybrid molecular beam epitaxy. Sm$^{3+}$ substitutes for Sr$^{2+}$, resulting in controlled electron doping of the films. Samarium was chosen as a dopant because of its high vapor pressure, but lanthanum-doped films showed similar behavior. A static in-plane compressive strain was produced by a 1\% lattice mismatch between the SrTiO$_3$ film and the substrate. Temperature-dependent longitudinal and Hall resistance measurements were performed using a Quantum Design Physical Property Measurement System. Hall measurements at 300 K were used to determine bulk carrier densities $n_\mathrm{3D}$. Further details of film growth and transport measurements are described in Ref.~\citenum{ahadi2019}. Optical SHG experiments were performed using an ultrafast laser supplying 40-fs pulses at a 10-kHz repetition rate and a center wavelength of 800~nm. The laser spot size was 30~$\mu$m and the fluence was below 20~mJ/cm$^2$. Data were collected over a wide temperature range from 10~K to 200~K in 5-K steps. The scattering plane angle $\phi$ was rotated through 360$^\circ$ to collect rotational anisotropy patterns at every temperature. Light polarization dependence was determined using in-plane (S$_\mathrm{in/out}$) and out-of-plane (P$_\mathrm{in/out}$) polarization geometries for incident and outgoing beams.

\section{Ferroelectric phase transition and origin of resistivity anomalies}

\begin{figure}[t]
\includegraphics{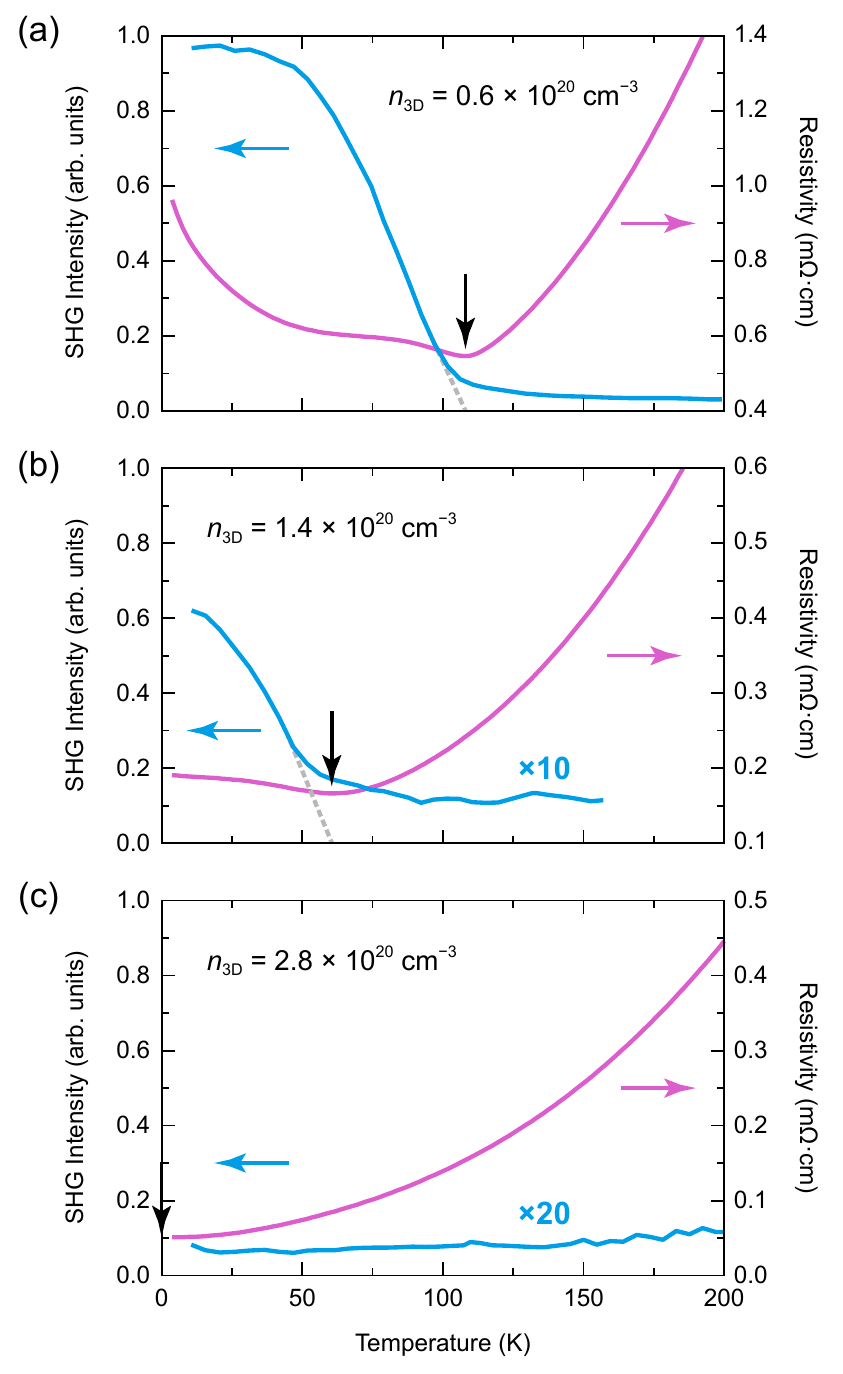}
\caption{\label{FigResistivity} Correlation between the onset of ferroelectricity and resistivity anomalies. Temperature dependence of SHG intensity (left axis) and electrical resistivity (right axis) for films with bulk carrier densities of (a)~$n_\mathrm{3D} = 0.6 \times 10^{20}$~cm$^{-3}$, (b)~$n_\mathrm{3D} = 1.4 \times 10^{20}$~cm$^{-3}$, and (c)~$n_\mathrm{3D} = 2.8 \times 10^{20}$~cm$^{-3}$. Black arrows mark local minima in the resistivity, which correlate with the onset temperatures of the ferroelectric order parameter.}
\end{figure}

Figure~\ref{FigResistivity} compares the temperature dependence of SHG intensity and electrical resistivity for three films with varying carrier densities. A small but finite SHG signal is measured at high temperatures for all films, but a sharp onset is observed at low temperatures for the $n_\mathrm{3D} = 0.6 \times 10^{20}$~cm$^{-3}$ and $1.4 \times 10^{20}$~cm$^{-3}$ films. Remarkably, local minimum anomalies in the resistivity occur at precisely the same temperature as the onset of SHG. We interpret these results as follows: At high temperatures, a nonzero SHG response exists due to inversion symmetry breaking by the dissimilar substrate and vacuum interfaces, producing a small polarizing field extending across the thickness of the film. As the temperature is reduced in the two lower-doped films, ferroelectric phase transitions occur resulting in a sharp increase in the SHG response. At the temperature where spontaneous ferroelectric order sets in, screening of the polar charges causes carriers to localize, decreasing the carrier density and increasing the film resistivity~\cite{raghavan2016}. In the highest doped film, the dipolar interactions necessary for spontaneous ferroelectricity are sufficiently screened by charge carriers, and long-range order is avoided. These observations demonstrate an unambiguous link between ferroelectricity and electrical transport signatures that have previously been associated with an enhancement of superconductivity~\cite{ahadi2019}. Our results not only settle the question of the origin of the resistivity anomalies, but in doing so confirm a direct connection between ferroelectricity and superconductivity. In addition, the results show that even at relatively high carrier densities, SrTiO$_3$ remains susceptible to a ferroelectric instability.

\section{Symmetry of the ferroelectric order parameter}

\begin{figure*}[t]
\includegraphics{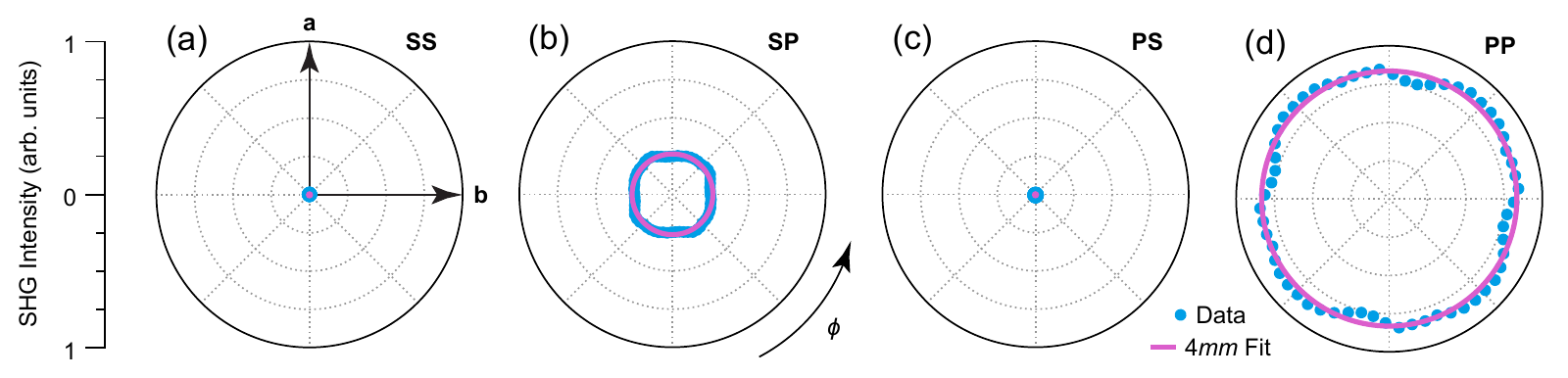}
\caption{\label{FigPetals} Symmetry properties of the ferroelectric phase. SHG rotational anisotropy patterns for (a)~S$_\mathrm{in}$-S$_\mathrm{out}$, (b)~S$_\mathrm{in}$-P$_\mathrm{out}$, (c)~P$_\mathrm{in}$-S$_\mathrm{out}$, and (d)~P$_\mathrm{in}$-P$_\mathrm{out}$ polarization geometries measured at 10~K on a $n_\mathrm{3D} = 0.6 \times 10^{20}$~cm$^{-3}$ film. Patterns have been four fold symmetrized to reduce noise. The polarization and angular dependence are consistent with a noncentrosymmetric $4mm$ point group, where SHG is forbidden for S$_\mathrm{out}$ geometries and $\phi$ independent for P$_\mathrm{out}$ geometries. $4mm$ is the isotropy subgroup of the $4/mmm$ high-temperature tetragonal point group generated by an $A_{2u}$ out-of-plane ferroelectric order parameter.}
\end{figure*}

To determine the symmetry of the ferroelectric order parameter, we performed SHG rotational anisotropy measurements at low temperatures. Such measurements can be used to precisely determine the full crystallographic and electronic point group symmetries of a material by measuring the individual elements of the nonlinear optical susceptibility tensor~\cite{harter2015,harter2017,zhao2018}. This tensor relates the polarization response of the material at the second harmonic to the incident electric field via the equation $P_i(2\omega) = \chi_{ijk}E_j(\omega)E_k(\omega)$. Figure~\ref{FigPetals} shows rotational anisotropy patterns for all four polarization channels: S$_\mathrm{in}$-S$_\mathrm{out}$, S$_\mathrm{in}$-P$_\mathrm{out}$, P$_\mathrm{in}$-S$_\mathrm{out}$, and P$_\mathrm{in}$-P$_\mathrm{out}$. We detect no measurable SHG in the S$_\mathrm{out}$ channels, while the P$_\mathrm{out}$ channels are independent of the scattering plane angle $\phi$. These observations are fully consistent with a $4mm$ ($C_{4v}$) polar point group, where $I_\mathrm{SS} = 0$, $I_\mathrm{SP} \propto \left|\chi_{zxx}\right|^2\sin^2\theta$, $I_\mathrm{PS} = 0$, and $I_\mathrm{PP} \propto \left|\chi_{zzz}\sin^2\theta - \chi_{zxx}\cos^2\theta\right|^2\sin^2\theta$ are predicted by symmetry considerations~\cite{boyd2008}. Here, $\theta \approx 30^\circ$ is the angle of incidence in the experiment. The relative magnitudes of the S$_\mathrm{in}$-P$_\mathrm{out}$ and P$_\mathrm{in}$-P$_\mathrm{out}$ SHG responses show that the dominant nonlinear susceptibility element is $\chi_{zzz}$, representing out-of-plane anharmonic motion of electrons driven by an out-of-plane electric field. The measured point group $4mm$ is an isotropy subgroup of the $4/mmm$ high-temperature tetragonal symmetry of the strained film. The symmetry lowering is generated by an $A_{2u}$ ferroelectric order parameter directed normal to the plane of the film, as expected for compressive strain~\cite{pertsev2000}. Identical symmetry is observed for films with other carrier densities and at other temperatures, including temperatures above the phase transition.

\section{Ising model of ferroelectricity}

\begin{figure}[hb]
\includegraphics{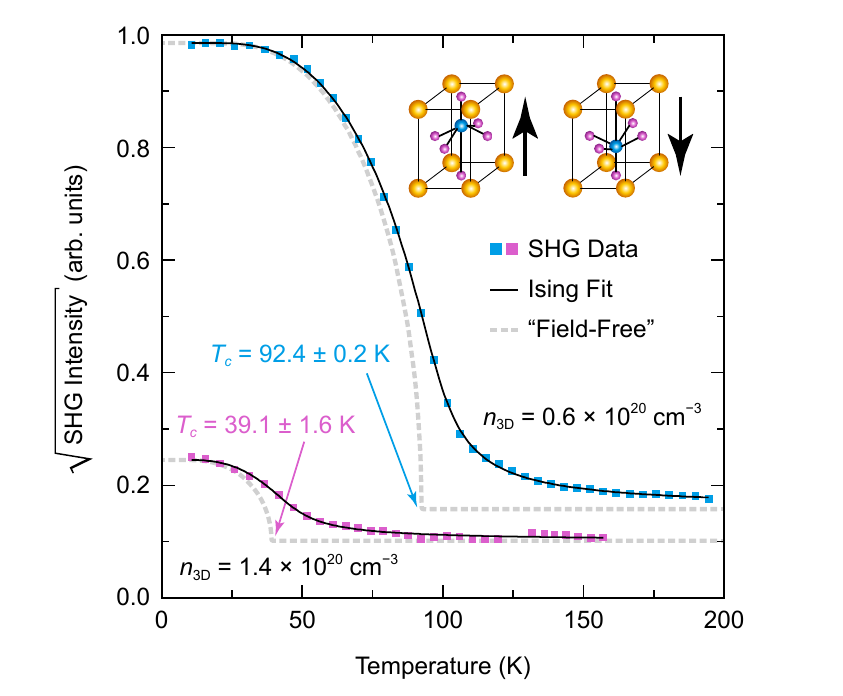}
\caption{\label{FigIsing} Ising model of the ferroelectric phase transition. The square root of SHG intensity, proportional to the ferroelectric order parameter, is reproduced exactly by a long-range mean-field Ising model that includes a small symmetry-breaking field arising from the substrate and vacuum interfaces of the thin film. The inset illustrates polar unit cell distortions equivalent to a bistable Ising order parameter. Dashed gray curves show the hypothetical order parameter in the absence of a field, from which the critical temperature can be estimated.}
\end{figure}

A long-range mean-field Ising model of ferroelectricity can be used to fit the temperature dependence of the SHG intensity, as shown in Fig.~\ref{FigIsing}. In this model, the two opposite polar unit cell distortions stabilized by in-plane compressive strain are equivalent to a bistable Ising order parameter (illustrated in the inset of Fig.~\ref{FigIsing}). In addition to a long-range electric dipole--dipole interaction term, a small extrinsic polarizing field breaking the symmetry of ``up'' and ``down'' configurations is included to account for the dissimilar substrate and vacuum environments sandwiching the film. The Hamiltonian of this model is
$${\mathcal{H} = -\frac{1}{2} \sum_{i,j} U(\textbf{r}_i - \textbf{r}_j) p_i p_j - E \sum_i p_i,}$$
where $i$ and $j$ label unit cells, $p_i = \pm 1$ is the Ising order parameter, $U(\textbf{r})$ is the dipolar interaction potential energy, and $E$ is the preexisting polarizing field. Because of the long-range nature of the interaction term, which ensures that a unit cell dipole will be influenced by many of its neighbors, a mean-field approximation is expected to be accurate. Within mean-field theory, the space-averaged order parameter $\left\langle p\right\rangle$ satisfies the self-consistency equation
$${\left\langle p\right\rangle = \tanh\left(\frac{\bar{U}\left\langle p\right\rangle + E}{k_B T}\right),}$$
where $\bar{U} = \sum_j U(\textbf{r}_j)/2 =  k_B T_c$ determines the critical temperature of the phase transition. $\bar{U}$ depends sensitively on carrier density because of electrostatic screening effects. To connect this statistical mechanical model to optical measurements, we assume $\chi_{ijk}$ to be proportional to the ferroelectric order parameter~\cite{harter2017} and include a small temperature-independent background term to account for secondary SHG from the film interfaces. Solutions to the self-consistency equation fitted to the data are plotted in Fig.~\ref{FigIsing}. From the fits, we extract bulk critical temperatures (phase transition temperatures without the polarizing field) of $T_c = 92.4 \pm 0.2$~K for $n_\mathrm{3D} = 0.6 \times 10^{20}$~cm$^{-3}$ and $T_c = 39.1 \pm 1.6$~K for $n_\mathrm{3D} = 1.4 \times 10^{20}$~cm$^{-3}$. Polarizing field energies are on the order of $E/k_B \sim 2$~K. Although the preceding model is simple, the goodness of the fits demonstrates that such a model can accurately capture the essential physics of ferroelectricity in strained SrTiO$_3$.

\section{Interplay between ferroelectricity and superconductivity}

\begin{figure}[t]
\includegraphics{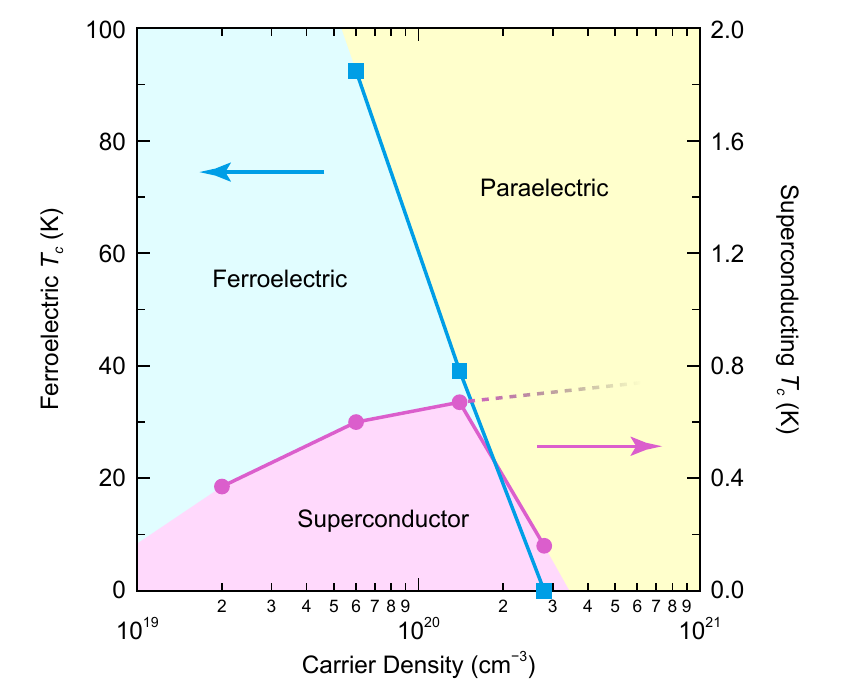}
\caption{\label{FigPhaseDiagram} Doping--temperature phase diagram of compressively strained SrTiO$_3$. In familiar analogy with other classes of unconventional superconductors, a ``dome'' of superconductivity exists in proximity to the quantum critical point of a coexisting order, in this case ferroelectricity. The dome appears to terminate abruptly at the point where long-range ferroelectric order is destroyed by screening from increased charge carriers, highlighting the essential role that ferroelectricity plays in the superconducting phase.}
\end{figure}

To place our observations in a broader context, we plot the measured doping--temperature phase diagram of compressively strained SrTiO$_3$ in Fig.~\ref{FigPhaseDiagram}, showing both the ferroelectric and superconducting phase boundaries (necessarily on different temperature scales). The values for the ferroelectric critical temperature are derived from the Ising model fits discussed above, which allow us to extrapolate thin film properties to a hypothetical bulk sample without an extrinsic polarizing field. The superconducting critical temperatures, measured in Ref.~\citenum{ahadi2019}, are for the same films studied in this work. The phase diagram shows the well-known superconducting ``dome''~\cite{collignon2019} in close proximity to the quantum critical point of the spontaneous ferroelectric order. In this region, the superconducting critical temperature is enhanced by up to a factor of two in comparison with identically doped strain-free films grown on SrTiO$_3$ substrates~\cite{ahadi2019}, highlighting the essential role played by the spontaneous ferroelectricity. We point out that the superconducting enhancement cannot be due to strain alone because the enhancement only occurs in a finite range of carrier densities near optimal $T_c$. Notably, the superconducting dome appears to terminate prematurely at the ferroelectric phase boundary, where screening by charge carriers overcomes any tendency towards long-range order. Taken together with the fact that the enhancement of $T_c$ occurs in the ferroelectric phase and away from the quantum critical point, this suggests that the ferroelectric order parameter itself may be playing a role in the superconductivity.

\section{Conclusions}

In conclusion, we used optical second harmonic generation to measure the doping and temperature dependence of the out-of-plane ferroelectric polarization in compressively strained SrTiO$_3$ thin films. We found a direct link between enhanced superconductivity and the onset of spontaneous ferroelectricity. Our results support a long-suspected connection between ferroelectricity and superconductivity in SrTiO$_3$ and should have significant value for developing and testing theoretical models relating the two phases. In particular, the success of the Ising model suggests a possible order-disorder component to the ferroelectric phase transition in strained SrTiO$_3$~\cite{blinc2003}. Furthermore, the observation of enhanced superconductivity deep within the ferroelectric phase where fluctuations are suppressed calls into question their role in superconducting pairing. Given our findings, exploring methods to stabilize ferroelectricity at higher doping levels via heterostructure engineering may prove fruitful. Another interesting direction to explore is enhancing spin-orbit coupling, which, in combination with ferroelectricity, may lead to exotic topological superconductivity~\cite{kozii2015,wang2016,kanasugi2018,kanasugi2019}.

\section*{Acknowledgments}

Optical measurements were supported by UCSB start up funds. Film growth and transport measurements were supported by the National Science Foundation (NSF) under Grants No.~1740213 and No.~1729489. L.D. acknowledges support by the Materials Research Science and Engineering Centers (MRSEC) program of the NSF under Grant No.~DMR-1720256 (Seed Program).

\end{document}